\documentclass[11pt]{article}

\usepackage{cite}
\usepackage{color}
\usepackage{amsfonts}
\usepackage{amsmath}
\usepackage{epsfig}
\usepackage{latexsym}
\usepackage{fancybox}
\usepackage{graphicx}
\usepackage{subfigure}
\usepackage{amssymb}

\numberwithin{equation}{section}
\allowdisplaybreaks

\def\spa#1{\phantom{\fbox{\rule[-#1cm]{0cm}{0cm}}}}

\newcommand{\beq}{\begin{equation}}

\newcommand{\eeq}{\end{equation}}
\newcommand{\bea}{\begin{eqnarray}}
\newcommand{\eea}{\end{eqnarray}}

\def\[#1\]{\begin{align}#1\end{align}}

\begin{document}

\hfuzz=100pt
\title{{\Large \bf{
Wormholes in 2d Hor\v ava-Lifshitz quantum gravity
}}}
\author{\sl Jan Ambjorn$^{a,b}$\footnote{ambjorn@nbi.dk}, \sl Yuki Hiraga$^{c}$\footnote{hiraga@eken.phys.nagoya-u.ac.jp}, 
\sl Yoshiyasu Ito$^{c}$\footnote{ito@eken.phys.nagoya-u.ac.jp} and \sl Yuki Sato$^{c,d}$\footnote{ysato@th.phys.nagoya-u.ac.jp}
  \spa{0.5} \\
\\
$^a${\small{\it The Niels Bohr Institute, Copenhagen University}}
\\ {\small{\it Blegdamsvej 17, DK-2100 Copenhagen, Denmark}}\\    
\\
$^b${\small{\it IMPP, Radboud University}}
\\ {\small{\it Heyendaalseweg 135, 6525 AJ, Nijmegen, The Netherlands}}\\    
\\
$^c${\small{\it Department of Physics, Nagoya University}}
\\ {\small{\it Chikusaku, Nagoya 464-8602, Japan}}\\
\\
$^d${\small{\it Institute for Advanced Research, Nagoya University}}
\\ {\small{\it Chikusaku, Nagoya 464-8602, Japan}}
\spa{0.3} 
}
\date{}

\maketitle
\centerline{}

\begin{abstract} 
We quantize the two-dimensional projectable Hor\v ava-Lifshitz gravity with a bi-local as well as space-like wormhole interaction.  
The resulting quantum Hamiltonian coincides with the one obtained through summing over all genus in the string field theory for two-dimensional causal dynamical triangulations.  
This implies that our wormhole interaction can be interpreted as a splitting or joining interaction of one-dimensional strings.    
      
\end{abstract}

\renewcommand{\thefootnote}{\arabic{footnote}}
\setcounter{footnote}{0}

\newpage

\section{Introduction}
\label{sec:introduction}

Quantum gravity in two dimensions is a pedagogical laboratory in which one can test ideas beyond the framework of perturbation theory. 
One of non-perturbative toy models for quantum gravity is two-dimensional causal dynamical triangulations ($2$d CDT) \cite{Ambjorn:1998xu}, 
in which one introduces a global time foliation to quantum geometries used in calculation of amplitudes of the theory.   
The $2$d CDT model regularizes a gravitational path-integral by a sum over geometries discretized by certain triangles which serve as a regulator,  and the continuum limit that removes the regulator can be performed analytically \cite{Ambjorn:1998xu}. 
Yet another example of $2$d quantum gravity that includes a global time foliation is the $2$d Hor\v ava-Lifshitz (HL) quantum gravity, 
which was originally introduced in higher dimensions to circumvent the perturbative non-renormalizability of quantum gravity \cite{horava}. 
It was shown that the quantum Hamiltonian of the $2$d projectable HL gravity becomes equivalent to the continuum Hamiltonian of $2$d CDT \cite{agsw} .
Projectable HL gravity is here defined to be the version of HL gravity where the lapse function that ensures time reparametrization invariance is a function of time only.

In the framework of $2$d CDT, a baby-universe creation is not allowed to occur due to the causality constraint imposed at the quantum level. 
In this regard, however, one can relax the constraint in such a way that quantum geometries keep the time-foliation structure, but baby universes can be created and annihilated, 
which is controlled by a ``string coupling constant''. This model is called the $2$d generalized CDT ($2$d GCDT) \cite{Ambjorn:2007jm}, 
and the sum of baby universes can be performed based on a third-quantization technique of string field theory for $2$d CDT \cite{Ambjorn:2008ta}, 
a continuum matrix model \cite{Ambjorn:2008jf} and a new scaling limit of matrix models \cite{Ambjorn:2008gk}. 
Therefore, topology of spacetime can change in $2$d GCDT, and in fact, the sum over all possible topologies, i.e. genus, can be done in the string field theory for $2$d CDT \cite{Ambjorn:2009fm}, 
which gives us non-perturbative multi-loop amplitudes. 
What is remarkable here is that such multi-loop amplitudes can be completely expressed in terms of a one-loop amplitude, 
and the one-loop amplitude, called the Hartle-Hawking wave function, is obtained as the zero-energy eigenstate of the continuum Hamiltonian of $2$d GCDT. 
Furthermore, the continuum Hamiltonian can be constructed by adding a simple interaction term into the continuum Hamiltonian of $2$d CDT, and it includes the effects from the sum over all genus\footnote{A related result is also known in 2d Liouville quantum gravity
\cite{moore,olga}.}.

$2$d CDT is born as a quantum theory, but by construction it does not allow for baby universes or wormholes. As mentioned 
we know how to generalize this quantum theory to include a summation over such configurations. 
Since the $2$d projectable HL quantum gravity is identical to $2$d CDT also this theory does not allow for baby universes or wormholes.
In this case we have a classical HL theory which can be canonically quantized (and results in $2$d CDT) \cite{agsw}. 
So it is natural to ask if 
there is a  generalized \textit{classical} $2$d HL theory which when canonically quantized will lead to the above mentioned GCDT 
quantum Hamiltonian which includes the summation over all baby universes and wormholes. It has been argued that in a low energy 
approximation of the quantum theory, the effect of wormholes and baby universes should be described by adding certain effective
bi-local terms to the effective action of the quantum theory (see for instance \cite{hawking} and references therein). The lack of a  higher dimensional theory of quantum gravity 
where one can perform detailed calculations has made these discussions qualitative rather than explicit. In this paper we show that 
in $2$d HL gravity one can add a term which is spatial  bi-local and that the corresponding theory can be canonically quantized and 
the resulting Hamiltonian is precisely the quantum Hamiltonian found in $2$d GCDT which includes the effects of baby universes and
wormholes. Canonical quantization does not allow in any straight forward way a change in the spatial topology, but we will argue
that the bi-local term we have added contains the seed for such a change.

This article is organized as follows: 
In section \ref{sec:HL}, we will review the $2$d projectable HL gravity, and explain its relation to $2$d CDT after the quantization. 
In addition, $2$d GCDT will be briefly explained. 
In section \ref{sec:HLwithwormhole}, we will introduce the wormhole interaction to the $2$d projectable HL gravity, explain how to quantize the system 
and show the relation between the $2$d projectable HL gravity with the wormhole interaction and $2$d GCDT. 
Section \ref{sec:discussion} is devoted to discussions.

\section{2d projectable HL gravity}
\label{sec:HL}
We briefly review the $2$d projectable Hor\v ava-Lifshitz (HL) gravity. 
Let us start with a $2$d spacetime manifold $\mathcal{M}$ equipped with a time foliation, i.e. 
\[
\mathcal{M} = \bigcup_{t \in [0,1]} \Sigma_t\ , 
\label{eq:foliation}
\]
where $\Sigma_t$ is a $1$d space labelled by $t$:
\[
\Sigma_t = \{ 
x^{\mu} \in \mathcal{M}\ | \
f(x^{\mu}) = t
\}\ , \ \ \ \text{with}\ \ \ \mu = 0,1\ .
\label{eq:embeddedhs}
\] 
Choosing that $f=x^0=t$, the time direction can be decomposed into the two directions, i.e. normal and tangential to $\Sigma_t$:
\[
(\partial_t)^{\mu} = \frac{\partial x^{\mu}}{\partial t} 
= N n^{\mu} + N^1E^{\mu}_1\ , 
\label{eq:timeevolution}
\]  
where $n^{\mu}$ and $E^{\mu}_1$ are respectively a unit normal vector and a tangent vector defined as 
\[
n^{\mu} = \left(1/N, -N^1/N \right)\ , \ \ \ E^{\mu}_1 = \delta^{\mu}_1\ . 
\label{eq:normaltangent}
\]
Here $N$ and $N^1$ are called the lapse function and the shift vector, respectively. 
Using eq.\ (\ref{eq:timeevolution}), a metric $g_{\mu \nu}$ on $\mathcal{M}$ has the Arnowitt-Deser-Misner form: 
\[
ds^2 = g_{\mu \nu}dx^{\mu}dx^{\nu} 
=-N^2 dt^2 
+h_{11} \left( dx +N^1 dt \right) \left( dx +N^1dt \right)\ , 
\label{eq:ADM}
\]
where $t=x^0$ and $x:=x^1$; $h_{11}$ is the spatial metric on $\Sigma_t$ defined as $h_{11} := E^{\mu}_1 E^{\nu}_{1}g_{\mu \nu}$. 

The $2$d HL gravity is a theory that keeps the structure of time foliation above, or in other words, 
it is invariant under the foliation preserving diffeomorphisms (FPD): 
\[
t \to t + \xi^0(t)\ , \ \ \ x \to x + \xi^1 (t,x)\ . 
\label{eq:fpd}
\]
Under the FPD (\ref{eq:fpd}), the fields transform as 
\[
\delta_{\xi} h_{11} 
&= \xi^0 \partial_0 h_{11} + \xi^1 \partial_1 h_{11} + 2 h_{11}\partial_1 \xi^1
\ , 
\label{eq:fpd1}
\\
\delta_{\xi} N_1 
&=
\xi^{\mu} \partial_{\mu} N_1 +N_1 \partial_{\mu} \xi^{\mu} +h_{11} \partial_0 \xi^1\ , 
\label{eq:fpd2}
\\
\delta_{\xi} N 
&= \xi^{\mu} \partial_{\mu} N + N \partial_0 \xi^0\ ,
\label{eq:fpd3}
\]  
where $N_1 = h_{11}N^1$. 

Note that if the lapse function $N$ is a function of time, $N=N(t)$, it stays as a function of time under the FPD. 
The $2$d projectable HL gravity satisfies this condition on $N$, and it is defined by the following action: 
\[
I =
\frac{1}{\kappa}
 \int dtdx\ N\sqrt{h}
\left(
(1-\lambda)K^2 -2\Lambda 
\right)\ ,
\label{eq:actionI}
\] 
where $\lambda$, $\Lambda$ and $\kappa$ are a dimensionless parameter, the cosmological constant and the dimensionless gravitational coupling constant, respectively; 
$h$ is the determinant of the metric $h_{11}$, i.e. $h=h_{11}$;  
$K$ is a trace of the extrinsic curvature $K_{11}$ defined as 
\[
K_{11} = \frac{1}{2N} 
\left( 
\partial_0 h_{11} 
-2\nabla_1 N_1
\right)\ , \ \ \ \text{with}\ \ \ \nabla_1 N_1 := \partial_1 N_1 - \Gamma^1_{11} N_1\ . 
\label{eq:extrinsiccurv}
\]
Here $\Gamma^1_{11}$ is the spatial Christoffel symbol:  
\[
\Gamma^1_{11} = \frac{1}{2} h^{11}\partial_1 h_{11}\ .  
\label{eq:Christoffel }
\]
In principle, one can add higher spatial derivative terms into the action (\ref{eq:actionI}), but they would not be important in $2$d and we omit such terms.

In \cite{agsw, Li:2014bla}, the quantization of $2$d projectable HL gravity was discussed, 
and in particular, it was shown that the quantum Hamiltonian coincides with the continuum Hamiltonian of $2$d CDT under the following identification of the parameters \cite{agsw}:
\[
\Lambda_{cdt} = \frac{\Lambda}{2(1-\lambda)}\ , \ \ \ \lambda < 1\ , \ \ \ \Lambda > 0\ ,
\label{eq:lambdacdt}
\]   
where $\Lambda_{cdt}$ is the renormalized cosmological constant in $2$d CDT\footnote{We have set unimportant dimensionless gravitational constant as $\kappa = 4(1-\lambda)$.}.

Essentially, the physics of the quantized system is described by the quantum mechanics of the spatial $1$d universe with the length invariant under the spatial diffeomorphism,  
\[
L (t) = \int dx \sqrt{h(t,x)}\ , 
\label{eq:L}
\]
with the Hamiltonian operator:
\[
\hat{H}^{(0)}_a(L) = (L\Pi^2)_a + \Lambda_{cdt} L\ , \ \ \ \text{with}\ \ \ a=0,\pm 1\ ,
\label{eq:CDThamiltonian}
\]
where the subscript $a$ specifies the operator ordering:  
\[
\left( 
L \Pi^2 
\right)_0 = - \frac{d}{dL} L \frac{d}{dL}\ , \ \ \ 
\left( 
L \Pi^2 
\right)_{-1} = - L \frac{d^2}{dL^2}\ , \ \ \ 
\left( 
L \Pi^2 
\right)_1 = - \frac{d^2}{dL^2}L\ .   
\label{eq:ordering}
\]
Each ordering has a precise geometric meaning (see \cite{agsw} for the detail) and makes $\hat{H}^{(0)}_a$ Hermitian in the product,  
\[
\langle \phi |\hat{H}^{(0)}_a| \psi \rangle 
= 
\int^{\infty}_{0}  \phi^* (L) \hat{H}^{(0)}_a \psi (L)\ d\mu_a (L) \ , 
\ \ \ 
\text{with}\ \ \ 
d\mu_a (L) := L^a dL\ ,
\label{eq:product1}
\]
where $\phi$ and $\psi$ are scalar functions. 
In this quantum theory, topology of the $1$d space stays the same. 

The $2$d generalized CDT ($2$d GCDT) allows for splitting and joining interactions of the $1$d space with the gauge-invariant length $L$, 
resulting in the change of topology. 
Based on the method of string field theory for $2$d CDT, in which the ``string'' means the $1$d universe, the summation over all possible genus has been done \cite{Ambjorn:2009fm}, 
and in that case, the quantum Hamiltonian (\ref{eq:CDThamiltonian}) is replaced with the following one:
\[
\hat{H}_a (L) = (L\Pi^2)_a + \Lambda_{cdt} L - G_s L^2\ , \ \ \ \text{with}\ \ \ a=0,\pm 1\ ,
\label{eq:GCDThamiltonian}
\]
where $G_s$ is a ``string coupling constant'' that controls the strength of splitting and joining interactions. 
The $L$ in (\ref{eq:GCDThamiltonian}) is still the length of a single connected spatial universe and it is remarkable that 
the effect of splitting such a universe in two or joining it with another universe and in the process over time 
changing the topology not only of the spatial universe but also changing the topology of spacetime in all possible ways,
seen from a single spatial universe is summarized by the simple last term  (\ref{eq:GCDThamiltonian}).
 Let us also remark that the so-called Hartle-Hawking wave function, again including the sum over all spacetime 
 topologies  is  the zero-energy eigenstate of the continuum Hamiltonian (\ref{eq:GCDThamiltonian}) \cite{Ambjorn:2009fm}.

In the next section, we will show that the continuum Hamiltonian of $2$d GCDT (\ref{eq:GCDThamiltonian}) can be reproduced 
if we quantize the $2$d projectable HL gravity with a wormhole interaction.

\section{Quantization of 2d projectable HL gravity with a wormhole interaction}
\label{sec:HLwithwormhole}

We consider the $2$d projectable HL gravity with a space-like wormhole interaction given by the action: 
\[
S &= \int dtdx\ \mathcal{L} \notag \\  
&= 
\frac{1}{\kappa}
 \int dtdx 
 N(t)\sqrt{h(t,x)}
\left(
(1-\lambda)K^2(t,x) -2\Lambda 
\right) 
 + \beta 
\int dt N(t) \int dx_1 dx_2 \sqrt{h(t,x_1)} \sqrt{h(t,x_2)}\ , 
\label{eq:shlw}
\]
where $\beta$ is a dimension-full coupling constant. 
One can show that the action (\ref{eq:shlw}) is invariant under the FPD (\ref{eq:fpd}) with the projectable lapse function, $N=N(t)$. 
The bi-local interaction in (\ref{eq:shlw}) relates two distinct points at an equal time.  
The action is a simplified version of the general bi-local action suggested in \cite{hawking}, made possible because HL 
gravity is invariant only under the foliation preserving diffeomorphisms    (\ref{eq:fpd}) and not the full set of diffeomorphisms.

Following the procedure used in \cite{agsw}, we wish to quantize the system governed by the action (\ref{eq:shlw}). 
We first introduce a new variable:
\[
\ell := \sqrt{h}\ . 
\label{eq:newvariables}
\]    
In terms of the variables (\ref{eq:newvariables}), 
the trace of extrinsic curvature is recast as  
\[
K = \frac{1}{N} 
\left( 
\frac{\partial_0 \ell}{\ell} 
- \frac{1}{\ell^2} \partial_1 N_1 
+  \frac{\partial_1 \ell}{\ell^3} N_1
\right)\ . 
\label{eq:newK}
\]
For passing to the Hamiltonian formalism, we introduce the conjugate momentum of $\ell$:
\[
\pi := \frac{\partial \mathcal{L}}{\partial \partial_0 \ell} = \frac{2(1-\lambda)}{\kappa} K\ . 
\label{eq:momentum}
\]
We then define the Hamiltonian $H$ and its density $\mathcal{H}$ through the Legendre transformation: 
\[
H
= \int dx \left( \pi \partial_0 l - \mathcal{L}  \right)
=: \int dx\ \mathcal{H} \ , 
\label{eq:H0}
\]
where 
\[
\mathcal{H}
= 
N_1 \left(- \frac{\partial_1 \pi}{\ell} \right) 
+ N
\left( 
\frac{\kappa}{4(1-\lambda)} \pi^2 \ell  
+ \frac{2}{\kappa} \Lambda \ell 
- \beta \ell \int dx_2 \ell(x_2)
\right)\ . 
\label{eq:densityH0}
\]
To obtain the Hamiltonian, we used 
\[
\pi \partial_0 \ell 
= \frac{\kappa}{2(1-\lambda)} N \pi^2 \ell  
- N_1 \frac{\partial_1 \pi}{\ell}
+ \partial_1 \left( N_1 \frac{\pi}{\ell} \right)\ , 
\label{eq:pildot}
\]
and omitted the last total derivative term. 
The variables, $\ell$ and $\pi$, satisfy the Poisson bracket: 
\[
\{ \ell (t,x),\pi (t,x') \} &= \delta (x-x')\ .  
\label{eq:PB}
\]

Since the lapse function and the shift vector are non-dynamical, we have the two kinds of constraint:
\[
\mathcal{C}^1 (t,x) &:= - \frac{\partial_1 \pi (t,x)}{\ell (t,x)} \approx 0\ , 
\label{eq:momentumconstraint}
\\  
\mathcal{C}(t) &:= \int dx 
\left( 
\frac{\kappa}{4(1-\lambda)} \pi^2(t,x) \ell (t,x) 
+ \frac{2}{\kappa} \Lambda \ell (t,x)
- \beta \ell (t,x) \int dx_2 \ell(t,x_2)
\right)
\approx 0
\ , 
\label{eq:constraints}
\]
where $\mathcal{C}^1(t,x)$ and $\mathcal{C}(t)$ are called the momentum constraint and the Hamiltonian constraint, respectively. 
Note that the Hamiltonian constraint becomes a global constraint due to the projectable condition, $N=N(t)$. 

Our strategy is to solve the local momentum constraint at the classical level, i.e.
\[
\mathcal{C}^1 (t,x) = 0\ ,  \ \ \ 
\Rightarrow 
\ \ \ \pi = \pi (t)
=: \Pi (t)\ ,
\label{eq:solvemomentumconstraint}
\]
and reduce the model to the $1$-dimensional system with the Hamiltonian, 
\[
H = N
\left(
\frac{\kappa}{4(1-\lambda)} \Pi^2 L
+ \frac{2}{\kappa}\Lambda L
-\beta L^2 
\right)
\ ,
\ \ \ 
\text{with} \ \ \ 
L(t) := \int dx\ \ell(t,x)\ ,
\]
where $L$ and $\Pi$ satisfy the Poisson bracket, $\{L(t),\Pi (t) \}=1$. 
Hereafter, without loss of generality, we choose 
\[
\kappa = 4 s_{\lambda} (1-\lambda)  > 0\ ,
\label{eq:kappa}   
\] 
where $s_{\lambda}$ is a signum function, $s_{\lambda} = \text{sgn} (1-\lambda)$. 
Accordingly, the Hamiltonian becomes
\[
H=N \left( s_{\lambda} 
\Pi^2 L + \tilde \Lambda  L
- \beta L^2
\right)
=N\mathcal{C}
\ , \ \ \ \text{with} \ \ \  
\tilde \Lambda = \frac{\Lambda}{2s_{\lambda} (1-\lambda)}\ .  
\label{eq:1dH}
\]

Let us first consider that $\beta = 0$, which is the case of the $2$d projectable HL gravity. 
As shown in \cite{agsw}, if $\Lambda/(\lambda -1 )\ge 0$, one can solve the Hamiltonian constraint $\mathcal{C}$ as 
\[
\Pi^2 = \frac{\Lambda}{2(\lambda - 1)}  \ge 0\ , 
\label{eq:c}
\]
resulting in a constant extrinsic curvature on the constraint surface. 
On the other hand, if $\Lambda/(\lambda -1 )<0$, 
the Hamiltonian constraint $\mathcal{C}$ yields $L(t)=0$. 
It is shown that if one performs a path-integral quantization, there is no formal difficulties associated with the quantization around $L(t)=0$, 
and this is the case that one can recover the continuum limit of $2$d CDT \cite{agsw}.

In the case that $\beta \ne 0$, the Hamiltonian constraint can be solved as 
\[
\Pi^2 = - \frac{\Lambda}{2(1-\lambda)} + s_{\lambda} \beta  L \ge 0 \ ,
\label{eq:c2}
\] 
for the following six cases:
\[
&(1)\ \ \ \beta > 0\ ,\ \Lambda > 0\ ,\ \lambda < 1\ , \ L \ge \tilde \Lambda/ \beta\ , \label{eq:1} \\
&(2)\ \ \ \beta > 0\ ,\ \Lambda > 0\ ,\ \lambda > 1\ , \ L \le \tilde \Lambda/ \beta\ , \label{eq:2} \\
&(3)\ \ \ \beta > 0\ ,\ \Lambda \le 0\ ,\ \lambda < 1\ , \ L \ge 0\ , \label{eq:3} \\
&(4)\ \ \ \beta < 0\ ,\ \Lambda \ge 0\ ,\ \lambda > 1\ , \ L \ge 0\ , \label{eq:4} \\
&(5)\ \ \ \beta < 0\ ,\ \Lambda \le 0\ ,\ \lambda < 1\ , \ L \le \tilde \Lambda/ \beta\ , \label{eq:5} \\
&(6)\ \ \ \beta < 0\ ,\ \Lambda < 0\ ,\ \lambda > 1\ , \ L \ge \tilde \Lambda/ \beta\ . \label{eq:6} 
\]
Otherwise, the constraint $\mathcal{C}$ requires $L(t)=0$ on the constraint surface. 
As in the $\beta=0$ case, when quantizing the system based on a path-integral, we don't have any problem with respect to the quantization around $L(t)=0$.

The classical $1$d system with the Hamiltonian (\ref{eq:1dH}) can be alternatively described by the following action: 
\[
S
= 
\int^{1}_{0} dt 
\left(
\frac{\dot L^2}{4s_{\lambda}NL} 
- \tilde \Lambda NL 
+ \beta NL^2
\right)\ ,
\label{eq:actionL}
\]
where $\dot L := dL/dt$. 
This system is invariant under the time reparametrization, $t \to t + \xi^0(t)$, which is ensured by the lapse function. 
In fact, the proper time, 
\[
T = \int^1_0 dt\ N(t)\ , 
\label{eq:propertime}
\] 
and the length, $L=L(t)$, are invariant under the time reparametrization, 
and so it makes sense to discuss the probability amplitude for a $1$d universe to propagates in the proper time $T(>0)$, 
starting from the state with the length $L_1$ and ending up in the one with  length $L_2$ \cite{agsw}. 
Such an amplitude can be computed based on a path-integral, and we evaluate it by a rotation to the Euclidean signature for convenience.  
In our foliated spacetime, for $\lambda < 1$, we can implement this procedure by a formal rotation, $t\to it$, which yields the amplitude:
\[
G(L_2,L_1;T) 
= \int \frac{\mathcal{D}N(t)}{\text{Diff}[0,1]} \int \mathcal{D}L(t) e^{-S_E [N(t),L(t)]}\ ,
\label{eq:cylinder}
\]
where $L(0)=L_1$ and $L(1)=L_2$; 
$\text{Diff}[0,1]$ is the volume of the time reparametrization; 
$S_E$ is the Euclidean action given by
\[
S_E
= 
\int^{1}_{0} dt 
\left(
\frac{\dot L^2}{4NL} 
+ \tilde \Lambda NL 
- \beta NL^2
\right)\ ,
\label{eq:actionE}
\] 
where $\dot L := dL/dt$. 
Hereafter, we will consider the case with $\lambda < 1$. 

We set $N=1$ as a gauge choice. One can show that the corresponding Faddeev-Popov determinant only gives an overall constant, 
which we will omit in the following. 
The amplitude (\ref{eq:cylinder}) then becomes
\[
G(L_2,L_1;T) 
= \int \mathcal{D}L(t) 
\exp \left[ 
- \int^{T}_{0} dt 
\left(
\frac{\dot L^2}{4L} 
+ \tilde \Lambda L 
- \beta L^2
\right)
\right]
\ ,
\label{eq:cylinder2}
\]  
which can be expressed in terms of the quantum Hamiltonian $\hat{H}$ that is unknown at the moment:
\[
G(L_2,L_1;T) 
=
\langle L_2 | e^{-T \hat{H}} | L_1\rangle\ ,
\label{eq:cylinder3}
\]  
where $| L \rangle$ is a quantum state of the $1$d universe with the length $L$. 
Based on a standard method (see e.g. \cite{agsw}), one can read off the quantum Hamiltonian from eq.\ (\ref{eq:cylinder2}) and eq.\ (\ref{eq:cylinder3}): 
\[
\hat{H}_a (L) = (L\Pi^2)_a + \tilde \Lambda L - \beta L^2\ , \ \ \ \text{with}\ \ \ a=0,\pm 1\ ,
\label{eq:GCDThamiltonian2}
\]
where $(L\Pi^2)_a$ is the same as the one defined in eq.\ (\ref{eq:ordering}), 
and accordingly the integral measure becomes 
\[
\mathcal{D}L(t) = \prod^{t=T}_{t=0} L^a(t)dL(t)\ . 
\label{eq:measure2}
\]   
The measure (\ref{eq:measure2}) gives a precise geometric meaning of the ordering  (\ref{eq:ordering}) \cite{agsw}.

As a result, if $G_s = \beta$ and $\Lambda_{cdt} = \tilde \Lambda = \Lambda/(2(1-\lambda))$ where 
$\beta>0$, $\Lambda > 0$ and $\lambda <1$,  
the quantum Hamiltonian (\ref{eq:GCDThamiltonian2}) is indeed equivalent to the continuum Hamiltonian of $2$d GCDT (\ref{eq:GCDThamiltonian}).

\section{Discussion}
\label{sec:discussion} 

We have quantized the $2$d projectable HL gravity with a space-like wormhole interaction. 
We have shown that the quantum Hamiltonian is equivalent to the continuum Hamiltonian of $2$d GCDT, 
 if $G_s = \beta$ and $\Lambda_{cdt} = \tilde \Lambda = \Lambda/(2(1-\lambda))$ where 
$\beta>0$, $\Lambda > 0$ and $\lambda <1$. 

In the parameter region corresponding to $2$d GCDT, let us consider the classical Hamiltonian constraint $\mathcal{C}$. 
When $\sqrt{\Lambda_{cdt}}L \ge 1/\xi$ where $\xi$ is a dimensionless quantity defined by $\xi := G_s/ \Lambda^{3/2}_{cdt}$, 
one can have on the constraint surface    
$
\Pi^2 = - \Lambda_{cdt} + G_s L \ge 0
$. 
However, when $\sqrt{\Lambda_{cdt}}L < 1/\xi$, the only allowed solution is $L=0$. 
In the case of the $2$d projectable HL gravity ($\beta = G_s = 0$), with the parameter region corresponding to $2$d CDT, i.e. $\Lambda_{cdt}>0$, 
the only solution to the classical Hamiltonian constraint is $L=0$. 
Therefore, when $\sqrt{\Lambda_{cdt}}L < 1/\xi$, the classical solution of the $2$d projectable HL with the wormhole interaction would be close to that of the $2$d projectable HL gravity, 
if one sits in the parameter region above; they can be quite different when $\sqrt{\Lambda_{cdt}}L \ge 1/\xi$. 
Such a relation also holds at the quantum level: As shown in \cite{asw}, the eigenfunctions of the continuum Hamiltonian $\hat{H}_{-1}$ of $2$d GCDT (\ref{eq:GCDThamiltonian}) 
can be well approximated by the eigenfunctions of the continuum Hamiltonian $\hat{H}^{(0)}_{-1}$ of $2$d CDT (\ref{eq:CDThamiltonian}), 
when $\sqrt{\Lambda_{cdt}}L < 1/\xi$. On the other hand, when $\sqrt{\Lambda_{cdt}}L \ge 1/\xi$, their behaviors are quite different, 
and in this case, in order for the theory to be well-defined, the unbounded nature of $\hat{H}_{-1}$ should be counteracted by the kinetic term, which would be a reflection of the classical Hamiltonian constraint (\ref{eq:c}). 

The picture of creation and annihilation of baby universes and wormholes   is conceptually straightforward in the string field
theory formulation of $2$d GCDT \cite{Ambjorn:2008ta}. Nevertheless  it is somewhat surprising that one from this can derive an effective 
Hamiltonian which can describe  propagation of a single spatial universe, i.e.\ the propagation where the spatial universe starts
with the topology of a circle and at a later time $T$ has the same topology, but where it in the intermediate times  is allowed
to split in two and either one part disappears in the vacuum (a baby universe), or the two parts join again at a later time (then 
changing the spacetime topology). This process of joining and splitting  can be iterated at  intermediate times 
and using string field theory we can perform the summation of all iterations and  derive the effective 
Hamiltonian (\ref{eq:GCDThamiltonian}). From the point of view of unitary evolution (or Euclidean time, semigroup evolution, to 
be more precise),  as given in eq.\ (\ref{eq:cylinder3}), it is difficult to understand how a complete set of intermediate states can 
both be given by the one spatial universe states $| L \rangle$ and by the complete multi-universe Fock states of the string field theory. 
However, this seems to be the case by explicit calculation, and we find it even more surprising that the simplest wormhole interaction term added to the classical action as in eq.\ (\ref{eq:shlw}) leads to precisely the same  single universe quantum Hamiltonian as found in the GCDT string field theory.

\section*{Acknowledgement}
The work of YS was supported by JSPS KAKENHI Grant Number 19K14705.




\end{document}